\documentclass[aps,pra,twocolumn,showpacs,superscriptaddress,preprintnumbers]{revtex4}

\usepackage{mathbbol}
\usepackage{bbm}
\usepackage[intlimits]{amsmath}
\usepackage{amssymb,amsthm} 
\usepackage{amsfonts}
\usepackage{latexsym}
\usepackage{float}

\usepackage{float}
\usepackage[ansinew]{inputenc}
\usepackage[dvips]{graphicx} 
\usepackage{epsfig}
\usepackage{ifthen}    
\usepackage{ae}
\usepackage{pst-all}
\usepackage{hyperref}

\newcommand{\trz}[2]{\, \mathrm{tr}_{#1}\! \left( #2 \right)}

\newcommand{\etal}{{\it et al.\ }}

\newcommand{\bra}[1]{\langle #1|}
\newcommand{\ket}[1]{| #1\rangle}

\newcommand{\braket}[2]{\left< #1\,\right|\left.\! #2\right>}

\newcommand{\ketbra}[2]{| {#1}\rangle \langle {#2} |}

\newcommand{\proj}[1]{\ketbra{#1}{#1}}
\newcommand{\bracket}[3]{\left< #1\right|#2\left| #3\right>}

\newcommand{\1}{\mathbbm{1}}

\renewcommand{\vr}{\varrho}
\newcommand{\vp}{\varphi}
\newcommand{\BE}{\begin{equation}}
\newcommand{\EE}{\end{equation}}
\newcommand{\be}{\begin{equation}}
\newcommand{\ee}{\end{equation}}
\newcommand{\bea}{\begin{eqnarray}}
\newcommand{\eea}{\end{eqnarray}}

\newcommand{\NN}{\ensuremath{\mathcal{N}}}

\newcommand{\WW}{\ensuremath{\mathcal{W}}}

\newcommand{\CC}{\ensuremath{\mathcal{C}}}
\newcommand{\cCC}{\ensuremath{{\rm conv}(\mathcal{C})}}

\renewcommand{\tt}[1]{\texttt{#1}}

\begin{document}

\title{An algorithm for characterizing SLOCC classes of
multiparticle entanglement}

\author{Hermann Kampermann}
\email{kampermann@thphy.uni-duesseldorf.de}
\affiliation{Institut f\"ur Theoretische Physik III, Heinrich-Heine-Universit\"at D\"usseldorf, D-40225
D\"usseldorf, Germany}

\author{Otfried G\"uhne}
\email{otfried.guehne@uni-siegen.de}
\affiliation{Naturwissenschaftlich-Technische Fakult\"at,
Universit\"at Siegen, Walter-Flex-Str. 3, D-57068 Siegen, Germany}

\author{Colin Wilmott}
\email{wilmott@fi.muni.cz}
\affiliation{Institut f\"ur Theoretische Physik III, Heinrich-Heine-Universit\"at D\"usseldorf, D-40225
D\"usseldorf, Germany}
\affiliation{Faculty of Informatics, Masaryk University, 60200 Brno, Czech Republic}

\author{Dagmar Bru\ss}
\email{bruss@thphy.uni-duesseldorf.de}
\affiliation{Institut f\"ur Theoretische Physik III, Heinrich-Heine-Universit\"at D\"usseldorf, D-40225
D\"usseldorf, Germany}

\date{\today}

\begin{abstract}
It is well known that the classification of pure multiparticle entangled states according
to stochastic local operations leads to a natural classification of mixed states in terms
of convex sets. We present a simple algorithmic procedure to prove that a quantum state lies
within a given convex set. Our algorithm generalizes a recent algorithm for proving separability
of quantum states [J. Barreiro {\it et al.}, Nature Phys. {\bf 6}, 943 (2010)]. We give several
examples which show the wide applicability of our approach.
We also propose a procedure to determine a vicinity of a given quantum state which still belongs to
the considered convex set.
\end{abstract}

\pacs{03.67.-a, 03.67.Mn, 03.65.Aa}
\maketitle

\section{Introduction}
The importance of quantum entanglement for quantum
computation as well as for many other applications in
quantum information processing has raised many fundamental
questions regarding its characterization \cite{EntanglementReviewHorodecki}.
In mathematical terms, a quantum state is said to be separable, if it can
be written as a mixture of projectors onto product states, otherwise it is entangled.
Much work has been devoted to the development of criteria (in particular, the well-known tool of entanglement witnesses) which can
prove that a quantum state is entangled, which means that it is {\it outside}
of the convex set of separable states \cite{EntanglementReviewOtfried}.
Interestingly, methods which prove that a mixed quantum state is {\it within}
the set of separable states (e.g., by providing an explicit decomposition into product states), are less well known.
Nevertheless, for some cases explicit decompositions are known \cite{WoottersConcPRL98,VidalDecomp98,Razmik,KayOptDetEntGHZPRA11}
and recently even some algorithms for this task have been developed \cite{spedalieri-2007-76,navascues-2009-103,ExpMultiPartEntDynNatPhys}.

If more than
two particles are considered, the problem becomes more complicated, since
different classes of multiparticle entanglement exist. One possibility
uses the notion of stochastic local operations and classical communication
(SLOCC) \cite{SLOCCClassesCiracPRA2000,SLOCCClassesVertraetePRA2002}. For this notion, one can again ask, whether
a given state can be decomposed into states of the same SLOCC class, which
leads for the  case of three spin-1/2 particles to the well known classification
into GHZ- and W-states \cite{08ClassThreeQubitEntBruss}. Distinguishing these classes is a hard task,
for partial results see Refs. ~\cite{08ClassThreeQubitEntBruss, JSiewertPrivComm}. Especially if one wishes to prove that
a given state is within an entanglement class (such as the W-class), no general methods are known.

In this paper we propose an algorithm which allows to prove that a given mixed state belongs to a specific SLOCC entanglement class i.e., a decomposition exists where the pure states belong to the specified entanglement class.
During the iterative procedure pure states and probabilities of the decomposition are determined as well as the ``rest'' operator which has an increasing mixedness during the iterations.
In case of convergence we determine an explicit decomposition of the initial mixed state in terms of a convex combination of projectors onto pure states with the desired properties as well as a ``rest'' which is verified to be fully separable. It is not known whether the algorithm converges in all cases, but the method is easy to implement and it turns out to be well working in practice. Our algorithm is a generalization of the algorithm for proving separability from Ref.~\cite{ExpMultiPartEntDynNatPhys}.

Additionally we present a simple method to determine a lower bound on the $\varepsilon$-ball of the calculated decomposition \cite{PermutationPolytopesPLA10}. This we achieve by constructing a cross polytope inside the convex set spanned by our decomposition. We find a ball with respect to the Hilbert-Schmidt distance for states in this convex space. This method helps to verify that such properties also hold in case of reasonable small ``experimental'' errors. In some cases it also helps to get an idea of how far the state is at least away from sets with different properties.

The paper is organized as follows. After an overview about multipartite entanglement and SLOCC entanglement classes in Section \ref{Sec-DefNot} we present in Section \ref{Sec-IdeaAlg} an iterative algorithmic procedure to incrementally determine constituents of a decomposition of a given mixed quantum state. The main ingredient, the maximization of the overlap of a pure state with a given density operator under SLOCC operations, will be presented in Section \ref{Sec-SLOCC} and the properties of the algorithm in terms of convergence as well as scaling is discussed in Sections \ref{sec-optepsilon}-\ref{Sec-ConvAlg}. It follows in Section \ref{Sec-Eball} the procedure of determining an $\varepsilon$-ball.
At the end a series of examples, where decompositions are determined and $\varepsilon$-balls are calculated, is presented in Section \ref{Sec-Examples} and compared with results already known in the literature. We conclude with possible further improvements and limitations of our procedure.

\section{Definitions and notations}\label{Sec-DefNot}
Let us first consider pure states. Generally, a pure composite quantum state of
$n$ particles is called entangled
if it cannot be written as a tensor product of local states
\begin{equation}
\ket{\psi_E}\neq\bigotimes_{i=1}^k\ket{\phi_i}.
\end{equation}
where the $\ket{\phi_i}$ are states on a subset of all $n$ particles.
Depending on the value of $k$ one can further distinguish the number
of parties involved which are not of tensor product form. Consequently,
the states are said to be biseparable ($k=2$), triseparable ($k=3$), up
to $n$-separable. If a state does not possess any tensor product
structure it is called genuine multipartite entangled.

This classification of pure state entanglement can be refined to
the equivalence under stochastic local operations and classical communication (SLOCC) \cite{SLOCCClassesCiracPRA2000,SLOCCClassesVertraetePRA2002}.
Physically speaking, SLOCC operations can be implemented
with nonzero probability via local operations and classical communication,  i.e.\ a single copy of $\ket{\psi}$ can be mapped onto $\ket{\phi}$ using local operations with probability $p>0$, but with probability $(1-p)$
some other state may result. If two states can be converted into each other via SLOCC, this implies that both states are in principle useful
for the same tasks in information processing, albeit the efficiency
might be different. We denote such a class of SLOCC equivalent states by $\mathcal{C}$.

Mathematically speaking, a general SLOCC operation can be represented by the action of local operators i.e., $A_{\mathrm{SLOCC}}=\bigotimes_i A_i$, where $A_i$ are arbitrary operators acting on the $i$th party. An SLOCC operation maps the initial state $\ket{\psi}$ to $\ket{\phi}$ by
\begin{equation}
\ket{\psi} \mapsto \ket{\phi}=\NN A_{\mathrm{SLOCC}}\ket{\psi},
\end{equation}
where $\NN$ denotes the normalization.

For mixed quantum states shared between $n$ parties a state is called entangled if it cannot be written as a convex combination of an $n$-fold tensor product of projectors onto pure states \cite{WernerEnt89} i.e.,
\begin{equation}
\varrho_{\mathrm{ent}}\neq \sum_j p_j \bigotimes_{i=1}^n\ketbra{\psi_{i}^{(j)}}{\psi_{i}^{(j)}}
\end{equation}
One can extend this naturally by considering $k$-separable states. Finally,  a state is genuine multipartite entangled, if it cannot be written as a mixture of biseparable states.

For three qubits and pure genuine entangled states there exist
two types of entanglement classes which are not SLOCC equivalent \cite{SLOCCClassesCiracPRA2000}: The two
representatives are the  GHZ state and the W state
\bea
\ket{GHZ}&=&\frac{1}{\sqrt{2}}(\ket{000}+\ket{111}),
\nonumber
\\
\ket{W}&=&\frac{1}{\sqrt{3}}(\ket{001}+\ket{010}+\ket{100}).
\label{wstate}
\eea
Any pure entangled state can either be transformed into $\ket{GHZ}$
or $\ket{W}$, but these two states cannot be converted into each other.
The analysis and the hierarchy of the set of mixed W/GHZ states was then
developed in Ref.~\cite{08ClassThreeQubitEntBruss}. For more than three
qubits already an infinite number of inequivalent SLOCC classes exist \cite{SLOCCClassesVertraetePRA2002}.

\section{The main idea for the algorithm}\label{Sec-IdeaAlg}

\subsection{Structure of the problem}
In this section, we will describe the main idea from Ref.~\cite{ExpMultiPartEntDynNatPhys}
to design an algorithm for proving that a quantum state $\vr$ belongs to a given class $\CC$. In the most general
case, the task is to prove that a quantum state $\vr$ is a convex combination of some
projectors onto pure states $\ket{\phi_k}$. This means that we can write
\be
\vr= \sum_k p_k \ketbra{\phi_k}{\phi_k},
\label{convdec}
\ee
where the $p_k$ form a probability distribution. The $\ket{\phi_k}\in \CC$ are states
within a specific class $\CC$: For instance, if one wishes to prove that $\vr$ is fully
separable, then $\CC$ is the class of pure fully product states, or if $\vr$ should be
proven to belong to the W-class for three qubits, then $\CC$ is the class of pure W class
states (that is, the SLOCC orbit of $\ket{W}$). In the following, we will denote the set
of density matrices that can be decomposed as in Eq.~(\ref{convdec}) as $\cCC$, meaning
the convex hull of $\CC$.

In order to design an algorithm to check whether  $\vr$ can be decomposed as in
Eq.~(\ref{convdec}) we will use the following two facts:

\underline{\it (a) Convexity:} First, the set of density matrices with a decomposition
as in Eq.~(\ref{convdec}) forms a convex set i.e., if $\vr_a$ and $\vr_b$
are in $\cCC$, then $(1-p)\vr_a + p\vr_b$ is also in $\cCC$. This is indeed
obvious by definition and it will be used as follows: Assume that we have
three states $\vr_a, \vr_b,$ and $\vr_c$ which obey
\be
\vr_b = \frac{1}{1-p}(\vr_a - p\vr_c)
\;\;
\Leftrightarrow
\;\;
\vr_a=(1-p)\vr_b + p\vr_c
\label{subtraction}
\ee
where  $0 \leq p \leq 1$ and $\vr_c$ lies in the set $\cCC$. In this situation, if we can
prove that $\vr_b \in \cCC$, then $\vr_a \in \cCC$ must hold, too. We will use this fact
in terms of the first part of Eq.~(\ref{subtraction}): starting from $\vr_a$ we will subtract
a term  $ p \vr_c$ with $\vr_c=\ketbra{\phi_k}{\phi_k}$ and $\ket{\phi_k}\in \CC$. If we can
then show that $\vr_b \in \cCC$, this implies that $\vr_a \in \cCC$.

\underline{\it (b) Highly mixed states are in $\cCC$:} As a second fact we need statements which imply
that highly mixed states are in $\cCC$. This, of course, requires a specification of the
degree of mixedness and depends on the structure of $\CC.$ For instance, if we consider
a bipartite $N\times M$ system and if $\CC$ denotes the set of pure product states, then it
has been shown that if
\be
\trz{}{\vr^2} \leq  \frac{1}{NM-1}
\label{cond2}
\ee
then $\vr$ is separable, that is  $\vr \in \cCC$ \cite{gurvits02}. Similar results have
been obtained for other situations \cite{zyczkowski-1998-58, PhysRevLett.83.1054, kendon-2002-49, 2006quant.ph..7195D}
For instance, an $N$-qubit state with $N \geq 3$
for which
\be
\label{fullsepbound}
\trz{}{\vr^2} \leq\frac{1}{2^N-\alpha^2}
\mbox{ with }
\alpha^2=\frac{2^N}{\tfrac{17}{2}3^{N-3}+1}
\ee
holds, is fully separable \cite{hildebrand-2006}. This condition will be used as a termination
condition in the algorithm below.

It should be noted that there are cases, where a condition as Eq.~(\ref{cond2}) is not directly
given. For instance, if $\CC$ is the set of all symmetric product states, then the maximally
mixed state is clearly not in $\cCC$, as $\cCC$ consists of matrices acting on the symmetric space only. Even for
the identity operator $\openone_S$ on the symmetric space, a condition like Eq.~(\ref{cond2}) is
not straightforward to derive, since there are non-symmetric states close to  $\openone_S$ which
are not in $\cCC.$ In our paper, such problems do not play a role, a detailed discussion of symmetric
states will be given elsewhere.

\subsection{The algorithm}
Now we can formulate the iterative algorithm to prove that a state $\vr$
is in $\cCC$. The algorithm consists of the following steps:

\begin{enumerate}

\item Take the input state $\vr$ as $\vr_k$ with $k=1.$

\item  Consider the optimization problem
\be
\max_{\ket{\phi}\in \CC} |\bra{\phi} \vr_k \ket{\phi}|
\label{step2}
\ee
and find some state $\ket{\phi_k}$ within $\CC$ which has
a high overlap with $\vr_k.$

The only aim is to find a state with high overlap, one does not need a certified
optimal solution of the maximization in Eq.~(\ref{step2}).  Also one may replace
in Eq.~(\ref{step2}) the matrix $\vr_k$ by  $\sqrt{\vr_k},$ which may improve
the convergence properties of the algorithm (see Section VII for more discussion).

\item
Find an $\varepsilon_k \geq 0$ such that
\be
\vr_{k+1} := \frac{1}{1-\varepsilon_k}( \vr_k - \varepsilon_k \ketbra{\phi_k}{\phi_k})
\label{step3}
\ee
has no negative eigenvalues and that furthermore
$\trz{}{\vr_k^2} \geq \trz{}{\vr_{k+1}^2}$ holds. In fact, one can directly calculate
the optimal $\varepsilon_k$, such that $\trz{}{\vr_{k+1}^2}$ is minimal (see
Section VI); this choice is, however, not mandatory. In practical implementations,
it can be useful to set an upper bound $\varepsilon_k \leq\varepsilon_{\rm max},$
as this prevents the algorithm from subtracting too much from $\vr_k$.

The main idea is that if $\ket{\phi_k}$ has a high overlap with $\vr_k$,
then it also has a high overlap with the eigenvector corresponding to
the maximal eigenvalue $\lambda_{\rm max}(\vr_k)$ of $\vr_k$. The
construction of $\vr_{k+1}$ leads typically to $\lambda_{\rm max}(\vr_{k+1}) \leq \lambda_{\rm max}(\vr_{k})$
and, due to the normalization, $\lambda_{\rm min}(\vr_{k+1}) \geq \lambda_{\rm min}(\vr_{k})$
holds. Hence, $\vr_{k+1}$ will be closer to the maximally mixed state
than $\vr_k$ and is more likely to obey conditions as in Eq.~(\ref{cond2}).

\item Check, whether $\vr_{k+1}$ fulfills conditions like Eqs.~(\ref{cond2}, \ref{fullsepbound}). If this is the
case, then $\vr_{k+1}$ is separable and due to Eq.~(\ref{subtraction}) also $\vr_k$
and finally $\vr_1$ are in $\cCC$. Then, the algorithm can terminate.

\item If $\vr_{k+1}$ does not fulfill Eqs.~(\ref{cond2}, \ref{fullsepbound})
return to step 2 and $k \mapsto k+1$ and iterate further until
Eqs.~(\ref{cond2}, \ref{fullsepbound}) hold for some $k$.

\end{enumerate}

Before discussing and extending this algorithm in detail, two facts must be mentioned:
First, it is of course not guaranteed that for a given state in $\cCC$ the algorithm
will terminate after a finite number of steps. So we do not claim that the algorithm
can in general prove that a state $\vr$ is in $\cCC$, we only claim (and demonstrate in this paper)
that the algorithm is a powerful tool which works very well in practice.

Second, a crucial step in the algorithm is the optimization in Eq.~(\ref{step2}). As
already mentioned, one does not need a certified solution, but still it is important
to find a good approximate solution. Clearly, the difficulty of this task
 depends on the  structure of $\CC.$

For the simple case that $\CC$ are the pure bipartite product states, one can do this
as follows: For the optimal $\ket{\phi}=\ket{a}\ket{b}$ the part $\ket{a}$ is the
eigenvector corresponding to the maximal eigenvalue of $X_A=\trz{B}{\vr_k \openone \otimes \ketbra{b}{b}}$ and $\ket{b}$
is similarly the vector corresponding to the maximal eigenvalue of
$X_B=\trz{A}{\vr_k \ketbra{a}{a} \otimes \openone}$. This can be used to
tackle the maximization iteratively: Starting from a random $\ket{a}$ one
computes the optimal $\ket{b}$ via $X_B$, then with this $\ket{b}$ the
optimal $\ket{a'}$, then again the optimal $\ket{b'}$ etc. In practice,
this converges quickly against the desired solution. For multiparticle fully
separable states, this can be done similarly \cite{guhne-2007-98}.

If $\CC$ denotes the SLOCC equivalence class of some pure state, however, it is not so
clear how to perform the optimization in  Eq.~(\ref{step2}). For pure three-qubit
W class states one may use the explicit parameterization of pure W states
from Ref.~\cite{08ClassThreeQubitEntBruss}, but for more qubits,
such explicit formulae are not available. A central step to extend the algorithm
from Ref.~\cite{ExpMultiPartEntDynNatPhys} to SLOCC classes is therefore a simple
algorithm for the maximization in Eq.~(\ref{step2}). Such an algorithm will be described in
the next section.

Finally, note that the termination conditions Eqs.~(\ref{cond2}, \ref{fullsepbound})
can also be used for SLOCC classes: From any pure state, one can obtain all pure product
states by (non-invertible) SLOCC, hence the fully separable mixed states are a subset of
$\cCC$.

\section{Maximizing the overlap of $n$-partite states via SLOCC}
\label{Sec-SLOCC}

As mentioned in the previous section, a crucial part of the algorithm is to
perform the maximization in Eq.~(\ref{step2}). If $\CC$ is the SLOCC orbit
of a suitably chosen $n$-partite pure quantum state $\ket{\Phi_0}$ in
$\mathcal H=\otimes_{i=1}^n \mathcal H_i$, the state after a general SLOCC
operation is given by
\begin{equation}
\ket{\Phi'}=\frac{\otimes_{i=1}^nA_i\ket{\Phi_0}}{\sqrt{\bracket{\Phi_0}{[\otimes_{i=1}^nA_i^\dag][\otimes_{i=1}^nA_i^{ }]}{\Phi_0}}},
\end{equation}
where $A_i$ is the local filtering (or SLOCC) operator of the $i$-th
party.

The goal is then to maximize the overlap of $\ket{\Phi'}$ with a given quantum
state $\vr$ by applying such an SLOCC operation, so one has to compute
\begin{equation}
\max_{\{A_i\}} \bracket{\Phi'}{\vr}{\Phi'}.
\end{equation}
The general optimization over the tensor product of SLOCC operators is a hard
task. Therefore, one may consider an iterative procedure where in each iteration
step ($\vp$) we optimize the overlap (fidelity) with respect to a single party
($i$) i.e., we calculate a new state by applying a local SLOCC operator of
the $i$-th party and using the identity for the remaining parties,
\begin{equation}
\label{Eq-phiNextStep}
\ket{\Phi_{\vp}}=
\frac{\1_{n\setminus i}\otimes A_{i_\vp}\ket{\Phi_{\vp-1}}}{\sqrt{\bracket{\Phi_{\vp-1}}{\1_{n\setminus i_\vp}^{ }\otimes A_{i_\vp}^{\dag } A_{i_\vp}^{ }}{\Phi_{\vp-1}}}},
\end{equation}
here $\1_{n\setminus i_\vp}$ denotes the identity operator on all parties except
the $i$-th party. In the following, we will usually omit  the symbol
$\1_{n\setminus i_\vp}$, when there is no risk of confusion.

In each iteration step the optimizing party is changed, e.g.\ by going from
the first to the second up to the $n$-th party and then starting with the first party again.
This iterative procedure is continued up to a fixed point, where the state
does not change anymore. The calculated SLOCC operator $A_{i_\vp}$ is
in this case up to some factor proportional to the identity. Note
that in general this optimization may converge to a local extremum only,
but, as discussed above, global optimality is not required for the separability
algorithm.

A possible way to deal with Eq.~(\ref{Eq-phiNextStep}) is to perform a direct numerical
optimization over the $A_{i_\vp}.$ If $d_{i}$ is the dimension of the local Hilbert space
$\mathcal H_i$ this requires an optimization over $2d_i^2-1$ real parameters. For
multi-qubit states this is directly feasible, however, for larger local dimensions or
for a large number of particles $n$, it is necessary to have an analytical method to
find an $A_{i_\vp}$ which increases the overlap. This analytical approach will
be explained in the following.

Increasing the overlap in each iteration step is equivalent to
(for brevity we use $i\equiv i_\vp$)
\be
\bracket{\Phi_\vp}{\varrho}{\Phi_\vp}
 =
\frac{\bracket{\Phi_{\vp-1}}{ A_i^{\dag}\vr A_i^{ }}{\Phi_{\vp-1}}}{\bracket{\Phi_{\vp-1}}{A_i^{\dag } A_i^{ }}{\Phi_{\vp-1}}}
\geq  \bracket{\Phi_{\vp-1}}{\varrho}{\Phi_{\vp-1}},
\label{Eq-Opti}
\ee
or equivalently
\begin{align}
 \nonumber
&\bracket{\Phi_{\vp-1}}{A_i^{\dag}\vr A_i^{ }}{\Phi_{\vp-1}}
\\
&- \bracket{\Phi_{\vp-1}}{\vr}{\Phi_{\vp-1}}\bracket{\Phi_{\vp-1}}{ A_i^{\dag } A_i^{ }}{\Phi_{\vp-1}}  \geq   0
\label{Eq-Opti1}.
\end{align}

In this inequality only the operator $A_i$ is unknown. We denote the overlap
(fidelity) of the previous iteration step by
$F_{\vp-1}=\bracket{\Phi_{\vp-1}}{\vr}{\Phi_{\vp-1}}$
and choosing a local orthonormal basis we can rewrite
\begin{eqnarray}
A_i & = & \sum_{r,s} a_{r,s} \ketbra{r}{s},\\
\varrho & = & \sum_{r,\xi_1,s,\xi_2}r_{r\xi_1,s\xi_2}\ketbra{r,\xi_1}{s,\xi_2} \;\; \mbox{and} \\
\ket{\Phi_{\vp-1}} & = & \sum_{r,\xi} c_{r\xi}\ket{r,\xi},
\end{eqnarray}
with $a_{r,s}\in \mathbbm{C}$, $r_{r\xi_1,s\xi_2}=r_{s\xi_2,r\xi_1}^*\in \mathbbm{C}$, and $c_{r\xi}\in \mathbbm{C}$, where we used the multi-indices $\xi_1,\xi_2$ which denote all index elements
of the $n\setminus i$-partite system.

With this parametrization the last term of the left hand side in Eq.~(\ref{Eq-Opti1})
takes the form
\begin{align}
&\bracket{\Phi_{\vp-1}}{A_i^{\dag } A_i^{ }}{\Phi_{\vp-1}}
=\trz{}{ A_i^{\dag } A_i^{ }\trz{n\setminus i}{\proj{\Phi_{\vp-1}}}}
\nonumber
\\
&  =\trz{}{ A_i^{\dag } A_i^{ }C}
=\sum_{h,j,l} a_{lh}^{*}a_{lj}^{ } (C)_{hj}
\nonumber
\\
&  =\sum_{h,j,l} a_{lh}^{*}a_{lj}^{ } (\tilde C)_{lh,lj}
 =\sum_{\zeta_1,\zeta_2} a_{\zeta_1}^{*}a_{\zeta_2}^{ } (\tilde C)_{\zeta_1,\zeta_2}
\nonumber
\\
&  = \bracket{a}{\tilde C}{a}.
\end{align}
Here, $C$ is just the matrix representation of $\trz{n\setminus i}{\proj{\Phi_{\vp-1}}}$
and we used $(\tilde C)_{l_1 h, l_2 j}=\delta_{l_1 l_2} C_{hj}$, that is
\begin{eqnarray}
(\tilde C)_{\zeta_1=(lh),\zeta_2=(lj)} & = & \left(\trz{n\setminus i}{\proj{\Phi_{\vp-1}}}\right)_{hj}\nonumber\\
(\tilde C)_{\zeta_1=(lh),\zeta_2=(i\neq l\, j)} & = & 0
\end{eqnarray}
and $\zeta_1,\zeta_2$ is a mapping of a two valued index
(the ``matrix element'' indices of the $i$-th party)
to a single valued integer index.

The remaining unknown expectation value of the left hand side in Eq.~(\ref{Eq-Opti1})
becomes
\begin{align}
&    \bracket{\Phi_{\vp-1}}{ A_i^{\dag}\varrho A_i^{ }}{\Phi_{\vp-1}}
  =  \sum_{{{{\scriptscriptstyle h,\xi_1}\atop {\scriptscriptstyle j,\xi_2}}\atop {\scriptscriptstyle l,m}}}
a_{hl}^* c_{l\xi_1}^* r_{h\xi_1,j\xi_2}^{ } a_{jm}^{ } c_{m\xi_2}
\nonumber
\\
&  =  \sum_{{{\scriptscriptstyle h,l}\atop {\scriptscriptstyle j,m}}}
a_{hl}^* a_{jm}^{ }\sum_{\xi_1,\xi_2}c_{l\xi_1}^* r_{h\xi_1,j\xi_2}^{ }c_{m\xi_2}
  =  \sum_{{{\scriptscriptstyle h,l}\atop {\scriptscriptstyle j,m}}}
a_{hl}^* a_{jm}^{ }D_{hl,jm}
\nonumber
\\
&  =  \sum_{\zeta_1,\zeta_2}a_{\zeta_1}^* D_{\zeta_1,\zeta_2} a_{\zeta_2}^{ }
  = \bracket{a}{D}{a}.
\end{align}
where
\be
  (D)_{\zeta_1=(hl),\zeta_2=(jm)}  =  \sum_{\xi_1,\xi_2}c_{l\xi_1}^* r_{h\xi_1,j\xi_2}^{ }c_{m\xi_2}.
\label{Eq-Dop}
\ee

Therefore, we can rewrite Eq.~(\ref{Eq-Opti1}) as
\begin{equation}\label{Eq-MaximizingEq}
\bracket{a}{D-F_{\vp-1}\tilde C}{a} \geq 0,
\end{equation}
where $F_{\vp-1}  =  \bracket{\Phi_{\vp-1}}{\varrho}{\Phi_{\vp-1}}$.
Note that the matrices $D$ and $\tilde C$ are hermitian and
$F_{\vp-1}$ is non-negative. The maximum left-hand side corresponds
to the maximal eigenvalue ($\lambda_{\mathrm{max}}$) of the matrix
$D-F_{\vp-1}\tilde C$ or likewise the left hand side of Eq.~(\ref{Eq-MaximizingEq})
is maximized by using the eigenvector corresponding to the maximal eigenvalue
($\ket{a_{\mathrm{max}}}$). By undoing the mapping we obtain the SLOCC operator,
\begin{equation}
\ket{a_{\mathrm{max}}}\rightarrow \tilde A_i.
\end{equation}
This procedure gives the following insight into the optimization over one local
filter as in Eq.~(\ref{Eq-phiNextStep}): If the maximal eigenvalue of $D-F_{\vp-1}\tilde C$
is positive, one can still increase the overlap with a suitable $A_i$.
The corresponding eigenvector gives an $\tilde A_i$ which increases the overlap.
Note, however, that this $\tilde A_i$ is optimal for Eq.~(\ref{Eq-Opti1}), but not
necessarily the optimal $A_i$ for $\bracket{\Phi_\vp}{\varrho}{\Phi_\vp}$ in Eq.\ \ref{Eq-Opti}.

In the practical implementation, especially at the beginning of the optimization
procedure it is not helpful to use $\tilde A_i$ directly as the SLOCC operator,
because this operator is not necessarily invertible. This could therefore
correspond to an irreversible operation which destroys entanglement.
It turns out that using
\begin{equation}
A_i=\tilde A_i+\lambda_{\mathrm{max}} \1_i
\end{equation}
avoids this problem. With this SLOCC operator the state for the next
iteration step is calculated according to Eq.~(\ref{Eq-phiNextStep}).
During the iteration procedure, $\lambda_{\mathrm{max}}$ will decrease
[see Eq.~(\ref{Eq-Opti1})], the SLOCC operators become close to the
identity, and the convergence criterion is that $\lambda_{\mathrm{max}}$
is up to numerical precision zero.

\section{Finding the optimal $\varepsilon_k$}
\label{sec-optepsilon}
A second optimization occurring in the algorithm is the task to find the best
$\varepsilon_k$ (see Eq.\ \ref{step3}). In detail, we want to maximize the decrease in the purity
in each iteration, that is
\begin{equation}
  \max_{\varepsilon_k} \left[\trz{}{\varrho_{k}^2}-\trz{}{\varrho_{k+1}^2}\right]
\end{equation}
with $\varrho_{k+1}=({\varrho_{k}-\varepsilon_k\proj{\phi_k}})/({1-\varepsilon_k})$.
With the abbreviation $\bracket{\phi_k}{\varrho_{k}}{\phi_k}=c$ the above maximization
leads to
\begin{equation}
  \max_{\varepsilon_k} \frac{\varepsilon_k^2\left[\trz{}{\varrho_{k}^2}-1\right]+2\varepsilon_k
\left[c-\trz{}{\varrho_{k}^2}\right]}
  {(1-\varepsilon_k)^2}.
\end{equation}
Taking the derivative with respect to $\varepsilon_k$, we find the maximum as
\begin{equation}\label{Eq-EpsMax}
  \varepsilon_{k}^\mathrm{max}=\frac{c-\trz{}{\varrho_{k}^2}}{1-c}.
\end{equation}
In the implementation, in case of $\varepsilon_{k}^{\mathrm{max}}
>10^{-2}\lambda_{d}$, we define
$\varepsilon_{k}^{\mathrm{max}}:= 10^{-2}\lambda_{d}$,
where $\lambda_{d}$ is the minimal eigenvalue of $\vr_{k}$ (guided by practical experience),
to keep the remaining state positive during the iterations. This corresponds
to an upper bound on $\varepsilon_k$, as mentioned above.

\section{Increasing the mixedness}
A central strategy of the algorithm is to increase the
``mixedness'' of $\varrho_{k}$ in each iteration step
i.e., to lower the purity $\trz{}{\varrho_{k}^2}.$ One may wonder,
whether this is always possible by subtracting some
$\ketbra{\phi_k}{\phi_k}\in \CC.$ Indeed, one can show
that this is the case, unless $\varrho_{k}$ is outside $\cCC$, which means that the algorithm has no
chance to succeed anyway.

To see this, the condition for an increase of the mixedness
can be formulated as
\begin{align}
\trz{}{\varrho_{k}^2} & >  \trz{}{\varrho_{k+1}^2}  =
\trz{}{\Big[\frac{\varrho_{k}-\varepsilon_k \proj{\phi_k}}{1-\varepsilon_k}\Big]^2}
\nonumber
\\
& =  \frac{1}{(1-\varepsilon_k )^2}
\left[\trz{}{\varrho_{k}^2}-2\varepsilon_k\trz{}{\varrho_{k}\proj{\phi_k}}+\varepsilon_k^2\right]
\nonumber\\
& \approx  \frac{1}{1-2\varepsilon_k}\left[\trz{}{\varrho_{k}^2}-
2\varepsilon_k\bracket{\phi_k}{\varrho_{k}}{\phi_k}\right]\nonumber
\end{align}
for small $\varepsilon_k$. It follows that iff
\begin{equation}
\label{Eq-maximizingOverlap}
\bracket{\phi_k}{\varrho_{k}}{\phi_k} > \trz{}{\varrho_{k}^2}.
\end{equation}
the state $\ketbra{\phi_k}{\phi_k}$ can be subtracted with a small weight,
and the mixedness increases.

If condition Eq.~(\ref{Eq-maximizingOverlap}) is not fulfilled for any $\ket{\phi_k}$
the mixedness cannot increase. But this implies that
\be
\sup_{\ket{\phi}\in \CC} \bracket{\phi}{\varrho_{k}}{\phi} \leq \trz{}{\varrho_{k}^2}.
\ee
Consequently, $\trz{}{\varrho_{k} \WW} \leq 0 $ for the observable
\be
\WW = \alpha  \openone - \varrho_k
\ee
with $\alpha = \sup_{\ket{\phi}\in \CC} \bracket{\phi}{\varrho_{k}}{\phi}.$ This $\WW$
is nothing but a witness \cite{EntanglementReviewOtfried} which discriminates between $\cCC$ and the remaining states,
and $\trz{}{\varrho \WW} < 0 $ implies that a state is not in $\cCC.$ Therefore,
states which cannot fulfill the condition in  Eq.~(\ref{Eq-maximizingOverlap}) are
either not in $\cCC$ or (in case that $\trz{}{\varrho_{k} \WW} = 0 $) they may lie
at the border of $\cCC.$ This, however, is a set of measure zero and not of
practical relevance.

\section{Convergence behavior of the algorithm}\label{Sec-ConvAlg}
Let us now discuss some practical issues. The question, whether or not
the algorithm converges depends first on the type of state and decomposition to be determined and second on the distance of the state from the boundary of the considered convex set. The closer the state is to a boundary the slower is the convergence. The algorithm does e.g.\ not work with rank deficit states, because overlap of the optimized pure states with the kernel of the density operator cannot be avoided i.e., it is not possible to ensure $\rho_{k+1}\geq 0$. In the three- and four-qubit case the decompositions consists of the order of $10^3$ states (meaning that the algorithm requires this number of iterations, until the conditions in Eqs.~(\ref{cond2}, \ref{fullsepbound}) apply) i.e., usually such decompositions contain many more states then the Caratheodory-bound of $d^2$.

In practice, the overlap optimization in Eq.~(\ref{step2}) with the
square root $\sqrt{\varrho}$ instead of  $\varrho$  has a better
convergence behavior. Also other fractional powers of $\varrho$ show a similar advantage. Note that replacing $\varrho$ by $\sqrt{\varrho}$
does not affect the proof that the iterated state is separable, if
Eqs.~(\ref{cond2}, \ref{fullsepbound}) apply.

\section{Lower bound on the $\varepsilon$-ball via cross polytope}
\label{Sec-Eball}

The presented algorithmic procedure allows to determine a decomposition of
the state $\vr$ with the specified SLOCC properties. After $n$ iterations we
have a decomposition i.e., the set   $\mathcal{S}=\{\{\proj{\phi_i}\},\vr_{n}\}$,
of our initial state of the form
\begin{equation}
\vr=\sum_{i=1}^n p_i\proj{\phi_i}+q_{ n}\vr_n,
\end{equation}
where the probabilities $p_i$ are given by $p_i=\varepsilon_{i}q_{i-1}$
and $q_{i}=\prod\limits_{j=1}^i \left(1-p_j\right)$ with $q_{0}:=1$ and $\vr_0:=\vr$.

By construction, our convex set $\mathcal{S}$ has specific ``entanglement'' properties
which are valid for all states in its convex hull. However, if we obtained the state $\vr$
from experimental data, we have to deal with errors and imperfections, and so the starting
state $\vr$ is affected by uncertainties. Therefore it is of great importance to give some
statements about the ``stability'' of the determined decomposition, or an estimate of the
probability that an experimental states lies inside this convex set.

A first possibility to deal with this problem was used in Ref.~\cite{ExpMultiPartEntDynNatPhys}: There, starting from
the experimentally obtained state $\vr_{exp}$ the measurements were simulated via a Monte-Carlo
simulation, and 200 sampled states were reconstructed via a maximum likelihood approximation.
Then, separability of the state $\vr_{exp}$ was only claimed, when the algorithm could prove that
$\vr_{exp}$ as well as all samples were separable. Note that the generation of states via Monte-Carlo
simulation of the measurements is a standard technique to estimate errors in ion-trap experiments.

A different possibility can be obtained by answering the question, how much can an experimental state
deviate from $\varrho$ such that the state still belongs to the set $\mathcal{S}$. Although this question
is in general not easy to answer \cite{ConvexPolytopes}, we can determine a lower bound on the minimal
Hilbert-Schmidt-distance of the state $\varrho$ with respect to the convex hull of $\mathcal{S}$.

\begin{figure}[t!!!]

\includegraphics{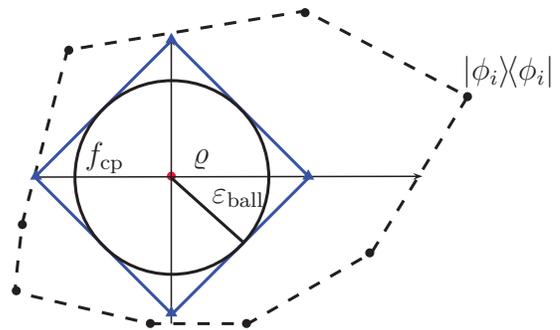} 

\caption{(Color online) Schematic two-dimensional example for a decomposition of $\varrho$ (red dot). The convex hull of the states of the decomposition i.e., the set $S$, is denoted by the dashed line. The cross polytope is shown in blue, the vertex states of the cross polytope are marked by blue triangles and the $\varepsilon_{\mathrm{ball}}$ corresponds to the circle.\label{Fig-CrossPoly}}
\end{figure}

The idea is to show that if the state deviates from $\vr$ in different directions, then it remains in
the convex set. More precisely, we construct a symmetric cross polytope \cite{PermutationPolytopesPLA10}
with the state $\varrho$ in the center (see Fig.\ \ref{Fig-CrossPoly} for a 2-dimensional example). For
a quantum state $\vr$ with Hilbert space dimension $d$ the set needs at least $d^2$ constituents s.t.
a nonzero volume object in this space is possible. The cross polytope is a symmetric polytope with
$2 (d^2-1)$ vertex states. The vertex states of the cross polytope are defined by
$\vec \varrho_{\pm i}^{\mathrm{cp}}=\vec \varrho\pm f_{\mathrm{cp}}\vec e_i$, where we used a vector
representation of the density operator in $(d^2-1)$-dimensional Euclidian space by mapping
\begin{eqnarray*}
&
A=
\begin{pmatrix}
a_1 & a_{d}+ia_{d+1} & \cdots & a_{3d-3}+i a_{3d-2} \\
\vdots & \vdots & \vdots & \vdots \\
 &  \cdots    & a_{d-1} & \cdots\\
 & \cdots      &  a_{d^2-2}-ia_{d^2-1} & 1-\sum_{i=1}^{d-1} a_i
\end{pmatrix}
\\
&
 \longrightarrow
\vec A=
\begin{pmatrix}
a_1 \\
a_2 \\
\vdots\\
a_{d^2-1}
\end{pmatrix}
.
\end{eqnarray*}
The elements of the basisvectors in this notation are given by $\left(\vec e_i\right)_{j}=\delta_{ij}$.

The Euclidian distance of the vertex states of the cross polytope with respect to the state $\varrho$, which is given by $f_{\mathrm{cp}}$, is maximized under the constraint that the vertex states are contained in the convex hull of our set $S$ (see Fig.\ \ref{Fig-CrossPoly}).
%

First, using a divide and conquer algorithm we calculate the maximal parameter $f$ of each state $\vec \varrho_{\pm i}=\vec \varrho\pm f\vec e_i$ such that it is contained in the convex hull of our set $S$. Whether a state is contained in the convex set can be decided by using a linear program, e.g.\ via the Matlab routine \tt{linprog}.
The vertex state with the smallest parameter $f_{\mathrm{cp}}$ is used for defining the vertex states of the cross polytope inside our convex hull.
The parameter $f_{\mathrm{cp}}$ also depends on the relative orientation of the chosen orthogonal basis $\{\vec e_i\}$, here an additional optimization is possible.

Then, the smallest Hilbert-Schmidt distance of $\varrho$ with respect to any point in the convex hull of the cross polytope is given by
$\varepsilon_{\mathrm{ball}}=\frac{f_{\mathrm{cp}}}{\sqrt{d^2-1}}$, where $d$ is the dimension of the Hilbert space and $f_{\mathrm{cp}}$ is the maximal parameter such that the cross polytope is contained inside the set $S$ \cite{PermutationPolytopesPLA10}. We will present an example below.

\section{Examples}\label{Sec-Examples}
In this section, we present several examples for the application of our algorithm outlined above.

\subsection{GHZ states affected by white noise}
First, we consider the GHZ state of $n$ qubits affected by white noise
\begin{equation}
\label{Eq-GHZWerner}
\varrho_{\mathrm{GHZ} n}(p)=p\proj{\mathrm{GHZ}_n}+\frac{1-p}{2^n}\1,
\end{equation}
where $\ket{\mathrm{GHZ}_n}=\frac{1}{\sqrt 2}\left(\ket{0\dots 0}+\ket{1\dots 1}\right).$
For three qubits, these states have the following properties:
\begin{itemize}
  \item $\varrho_{\mathrm{GHZ} 3}$ is fully separable iff $p\leq 1/5= 0.2$ \cite{WStatesDurCiracPRA2000},
  \item $\varrho_{\mathrm{GHZ} 3}$ is genuine multipartite (tripartite) entangled iff $p>3/7 \approx 0.4286$ \cite{GuehneEntNJP10},
  \item $\varrho_{\mathrm{GHZ} 3}$ belongs to the GHZ class iff $p \gtrsim 0.6955$. For a detailed discussion see
Ref. \cite{JSiewertPrivComm}.
\end{itemize}
With our algorithm we can determine separable and biseparable decompositions up to the threshold values of
$p$. A bound with a threshold value of $p=\frac{559}{805} \approx 0.6944 $ for existing W-class decomposition
was obtained with our algorithm, which is close to the optimal threshold value of $p\approx 0.6955$.

\subsubsection{The $\varepsilon$-ball and the robustness depending on $p$}
\begin{figure}[t!!]
\includegraphics[width=8cm]{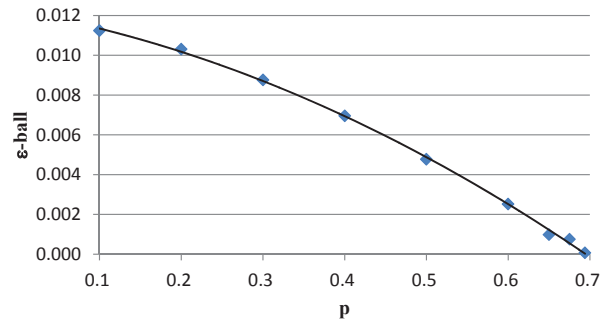}
\caption{(Color online) The lower bound on the size of the $\varepsilon$-ball of
the state given in Eq.\ (\ref{Eq-GHZWerner}) as a function of $p$. The
decompositions are obtained with the procedure of Sec.\ \ref{Sec-Eball}. The line is a polynomial fit of the points for guiding the eye. The ball-size strongly depends on the number of constituents in the decomposition, especially if states far away from the border are considered. Therefore here decompositions are compared which all consists of about 4000 pure states.    \label{Fig-EBall}}
\end{figure}

With our algorithm it is possible to obtain W-decompositions of the state $\rho_{\mathrm{GHZ} 3}$ up to $p=0.6944$. Due to
the ``small'' number of vertex states forming the convex set the size of an $\varepsilon$-ball generated
via the procedure of  Sec.\ \ref{Sec-Eball} will be small in comparison to the maximum possible ball which
fulfills the considered properties. However the lower bound on the size of this ball will also depend on the distance
of the considered state to the border where the properties are not fulfilled anymore.
In Fig.\ \ref{Fig-EBall} the lower bound on the size of the $\varepsilon$-ball is plotted versus the
parameter $p$. As expected for the almost maximally mixed state (small $p$) the
ball is quite large, but it decreases by several orders of magnitude as $p$ approaches
the threshold value of the considered convex set.
%

From our  $\varepsilon$-ball we can also deduce that any state $\varrho'$ which has an Euclidean distance from $\vr$ which is smaller equal the radius of the $\varepsilon$-ball is contained in our convex set i.e., using the Hilbert-Schmidt-distance between two operators $A$ and $B$ given by $d=\sqrt{\trz{}{(A-B)^2}}$, we can calculate the distance between two GHZ-Werner states, see Eq.\ (\ref{Eq-GHZWerner}), with purity $p$ and $p'=p+\delta$ to be given by
\begin{eqnarray}
  d^2 & = & \trz{}{[\varrho_{\mathrm{GHZ} }(p)-\varrho_{\mathrm{GHZ} }(p')]^2}\nonumber\\
   & = & \frac{7}{8}\delta^2.
\end{eqnarray}
If we have an $\varepsilon$-ball in the surrounding of $\varrho_{\mathrm{GHZ} }(p)$ with radius $\varepsilon_{\mathrm{ball}}$ then we can deduce that also the state $\varrho_{\mathrm{GHZ} }(p')$ with
\begin{equation}
\delta\leq \sqrt{\frac{8}{7}}\varepsilon_{\mathrm{ball}}
\end{equation}
belongs to the same convex set, allowing to increase the threshold parameter accordingly.

\subsubsection{Entanglement properties of $\rho_{\mathrm{GHZ} 4}(p)$}
The maximal parameter $p_{\mathrm{opt}}\approx 0.467$ for biseparability of
the four-qubit GHZ-state mixed with white noise was derived in Ref.~\cite{GuehneEntNJP10}.
Our algorithm is able to determine decompositions for values of $p$ up to $p\approx 0.466$.

For four qubits there is already a continuous set of inequivalent SLOCC entanglement classes,
and it can happen that a state can be decomposed into biseparable states, but not into SLOCC equivalents
of some genuine multipartite entangled states.  To investigate this, we considered the state
$\rho_{\mathrm{GHZ} 4}(p)$ and asked when it can be decomposed into SLOCC equivalents of the four-qubit W-state, $\ket{W_4}=\tfrac{1}{2}(\ket{0001}+\ket{0010}+\ket{0100}+\ket{1000})$.
This seems to be only possible for $p\leq 0.32$, but for the larger parameter regime  $p\leq 0.467$ the state
is biseparable.

An intuitive argument for such a behavior is that the four qubit W-state can be transformed via SLOCC
into a three-qubit GHZ state as well as to a two-qubit Bell state tensored with a product state, but
it is {\it not} possible to reach $\ket{\phi_4}=\ket{\phi^+}\otimes\ket{\phi^+}$, we even have $\max_{W\in\mathrm{SLOCC}(W_4)}|{\braket{W}{\phi_4}}|^2=0.5$\footnote{This property can e.g.\
be supported by considering the Schmidt-rank of
$\ket{W_4}$ with respect to the split 1,3 vs 2,4 which is 2, whereas the corresponding Schmidt rank of
$\ket{\phi^+}\otimes\ket{\phi^+}$ is 4. It is not possible to increase the Schmidt-rank of a state via SLOCC operations.}.
States like $\ket{\phi_4}$, however, are essential
in the biseparable decomposition of $\rho_{\mathrm{GHZ} 4}(p)$ \cite{GuehneEntNJP10}.
Many lower entangled pure states are SLOCC inequivalent to specific genuine multipartite entangled
pure states.

\subsection{W-states with white noise}
In order to give an example where the algorithm is not capable of computing the
threshold of separability, we consider states of the form
\be
\varrho_{Wn}(p)=p\proj{W_n}+\frac{1-p}{2^n}\1,
\ee
where the three-qubits W state $\ket{W_3}$ is given in
Eq.\ (\ref{wstate})
and for four qubits we have
$\ket{W_4}=\tfrac{1}{2}(\ket{0001}+\ket{0010}+\ket{0100}+\ket{1000})$.
The border to the class of biseparable states can only be roughly approached by the algorithm.
In the three-qubit case we have a gap of about $\Delta p=0.03$, the exact value $p=0.4790$
is known from Ref.~\cite{TamMultPartEnt}. For the four-qubit W-state
with white noise the gap with respect to the upper bound obtained by an
semidefinit-programming-witness (SDP-witness) \cite{TamMultPartEnt} is approximately $\Delta p_4=0.04$
i.e., we determined a decomposition for $p=1-0.526-0.04=0.434$.

\subsection{Bound entangled state from an unextendible product basis}\label{Sec-UPB}
As a second example, we consider the bound
entangled states arising from an unextendible product basis \cite{BennettUPB}. These states
are defined via using the product vectors $\ket{\psi_0}=\ket{0}(\ket{0}-\ket 1 )/\sqrt{2}$, $\ket{\psi_1}=(\ket{0}-\ket 1 )\ket{2}/\sqrt{2}$,
$\ket{\psi_2}=\ket{2}(\ket{1}-\ket 2 )/\sqrt{2}$, $\ket{\psi_3}=(\ket{1}-\ket 2 )\ket{0}/\sqrt{2}$,
$\ket{\psi_4}=(\ket{0}+\ket 1 +\ket 2 )(\ket 0+\ket 1+\ket 2)/3$.
Then the state
\begin{equation}
\varrho_{\mathrm{BE}}=\frac{1}{4}\left(\1 -\sum_{i=0}^4\proj{\psi_i}\right)
\end{equation}
is an entangled state on a $3\times 3$ system, which is not detected by the PPT
criterion.
We considered the family of states
\begin{equation}
\varrho_{\mathrm{UPB}}(p) = p\varrho_{\mathrm{BE}} +(1-p)\1/9.
\end{equation}
They have often been used as a test-bed for separability criteria. To our knowledge, the
best criterion for these states is the first step of the algorithm of Doherty et al.\
\cite{DohertyPRL02} which detects them to be entangled for $p> 0.8691$. Our algorithm proves that these states
are separable for $p\leq 0.83$.

\subsection{Thermal states with the Heisenberg interaction}\label{Sec-Heisenberg}
Let us consider the
thermal state
\begin{equation}
\varrho_{\mathrm{H}}(T)\sim \exp \{-H_H/T\}
\end{equation}
of three spin-1/2 particles interacting with
the Heisenberg interaction,
\begin{equation}
  H_H=\sum_{i<j}h_{ij}\;\; \mathrm{with}\;\; h_{ij}=\sum_{k=x,y,z}\sigma_k^{(i)}\otimes\sigma_k^{(j)},
\end{equation}
where $i,j\in\{1,2,3\}$.
In Ref.\ \cite{SpinSqueezeIneqToth} the entanglement properties of this system were investigated, and
it was shown that the spin-squeezing inequality $(\Delta J_x)^2 + (\Delta J_y)^2 +(\Delta J_z)^2\geq N/2$ with $J_k = 1/2\sum_i\sigma_k^{(i)}$ detects these states as entangled for $T\leq 5.461$.
Remarkably, the spin-squeezing inequality shows that for $4.329\leq T\leq 5.461$
the thermal state is biseparable with respect to any bipartition, but not fully
separable. Direct application of our algorithm gives that for $T\geq 5.462$
the thermal state is fully separable, giving strong evidence that the spin-squeezing
inequality is a necessary and sufficient criterion for the thermal state. For more
than three spins, however, this does not seem to be the case.

\subsection{Experimental pseudo bound entangled state}
\begin{figure}[t!]
\begin{center}
\includegraphics[width=9cm]{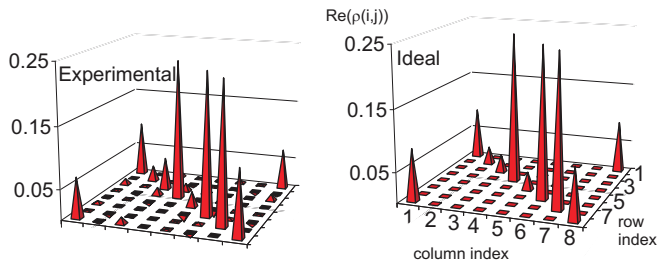}
\end{center}
\caption{(Color online) Experimental pseudo bound entangled state \cite{NMRBE}. All imaginary elements of the experimental state are small and therefore not shown. \label{Fig-ExpBEState}}
\end{figure}

In Ref.~\cite{08ClassThreeQubitEntBruss} a class of three-qubit bound entangled states with rank seven
were introduced. We consider the specific state where the entanglement is maximally robust with
respect to white noise \cite{08HyllusBEGenOWittnessPRA04}:
\begin{equation}\label{Eq-rhoBE}
\begin{split}
&\rho_{\mathrm{BE3}} =
\frac{1}{N}
\bigg(2 \proj{\mathrm{GHZ}}+\\
&
a\proj{001}+a\proj{010}+\frac{1}{a}\proj{011}+\\
&
\left. a\proj{100}+\frac{1}{a}\proj{101}+\frac{1}{a}\proj{110}\right),
\end{split}
\end{equation}
where
$\ket{\mathrm{GHZ}}=\frac{1}{\sqrt 2}\left(\ket{000}+\ket{111}\right)$, $a=0.3460$,
and the normalization is $N = \left(2+3( a+\frac{1}{a})\right)$.
The state has the curious property that it is biseparable with respect to any bipartite splitting, but it is nevertheless entangled.
A pseudo bound entangled state of this form was experimentally generated and characterized in Ref.~\cite{NMRBE}, see Fig.\ \ref{Fig-ExpBEState}. It was shown via a witness operator that the state is entangled and PPT with respect to any bipartite splitting. Now with our algorithm we are able to prove that this state is biseparable with respect to the split B-AC with an $\varepsilon$-ball of $\varepsilon_{\mathrm{ball}}=4\cdot 10^{-4}$.

%

\subsection{Experimental three-qubit W-state}
\begin{figure}[t!]
\begin{center}
\includegraphics[width=9cm]{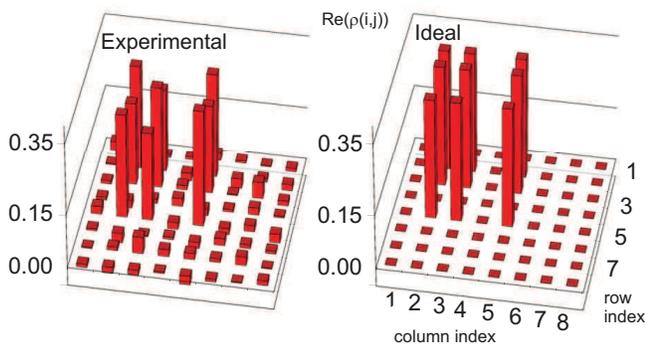}
\end{center}
\caption{(Color online) Real part of the experimental W-state from Roos et al. \cite{RoosPrivCom} (left) in comparison to the ideal W-state (right). All imaginary elements of the experimental state are small and therefore not shown.\label{Fig-ExpWState}
}
\end{figure}

Nowadays it is possible in several experimental setups to generate quantum states which are e.g.\ close to the three-qubit W-state.
The generated states are characterized by quantum state tomography. Here we use data from an experiment by Roos \etal \cite{RoosPrivCom}. This ``typical'' experiment (see Fig.\ \ref{Fig-ExpWState}) led to a fidelity of
$F=\sqrt{\bracket{W}{\varrho_{\mathrm{exp}}}{W}}=0.9$, where we can prove via an entanglement witness that this state is genuine multipartite entangled. Now the question arises whether this state really belongs to the W-class of entanglement.
With our algorithm it is not possible to find a W-class decomposition of $\varrho_{\mathrm{exp}}$, so probably the state belongs to the class of GHZ-entangled-states. On the other hand we can find for the slightly depolarized version of this state, e.g.\ $\varrho'=0.96\varrho_{\mathrm{exp}}+\frac{0.04}{8}\1$, a W-decomposition and we can prove that it is genuine multipartite entangled.

%

\subsection{Summary of the algorithmic performance}
In this section we summarize the obtained threshold parameters
for the various examples considered. Our standard approach was to consider different types of quantum states mixed with white noise of the form:
\begin{equation}\label{Eq-WhNoiseP}
  \vr_i(p)=p\vr_i+(1-p)\1/d.
\end{equation}
Then we determined the threshold parameters $p$ for various types of entanglement which we can achieve via our algorithm, and to compared them
with bounds from the literature or with exact values if they are known.
The only exceptional parameterization was the case of the thermal equilibrium Heisenberg spin chain state (Sec.\ \ref{Sec-Heisenberg}) which depends on the temperature.
In Table \ref{Tab-Decomp} we summarize the threshold parameters reached via our algorithm and show the best known bounds from the literature.
For several cases we also calculated  lower bounds on the $\varepsilon$-ball for our determined decompositions (see Sec.\ \ref{Sec-Eball}).
For all cases where the exact bounds are known we can reproduce the threshold parameters quite well by our procedure, apart from the three-qubit W state, where there is a small gap.
It is remarkable that our algorithm seems to work independently of the type of decomposition to be determined. For the given examples the calculation needs less than 5 min on a standard personal computer. The number of terms calculated in the decomposition strongly depends on the distance to the border of the convex set. I.e. already a very small change on the $p$-threshold values has a huge influence on the number of terms in the decomposition. For threshold values with the given precision in Table \ref{Tab-Decomp} it is possible to find decomposition within the order of $\sim 5000$ terms.

\begin{table}[]
\vspace{0.5cm}
\begin{center}
\begin{tabular}{|l|c|l|l|l|}
  \hline
  State & Ent. & Bound & Decomp. & $\varepsilon$-ball \\
  \hline \hline
$\vr_{\mathrm{GHZ} 3}$   & S     & $1/5^*$ (a)   & $0.199$   & $8.1\cdot 10^{-5}$ \\
        & BS    & $0.429^*$ \cite{TamMultPartEnt}  & $0.4285$  & $9.2\cdot 10^{-6}$ \\
        & W     & $0.6955^*$ \cite{JSiewertPrivComm}  & $0.694$   & $7.1\cdot 10^{-5}$ \\ \hline
$\vr_{\mathrm{GHZ} 4}$   & S     & $1/9^*$ (a)   & $0.111$   &  \\
        & BS    & $0.467^*$ \cite{TamMultPartEnt}   & $0.466$   &  \\
        & W     &        & $0.316$   &  \\ \hline
$\vr_{\mathrm{W3}}$     & S     & $ 3/11$ (a)   & $0.1727$  & $7.6\cdot 10^{-4}$ \\
        & BS    & $ 0.479^*$ \cite{TamMultPartEnt}  & $0.45$    & $1.1\cdot 10^{-3}$ \\ \hline
$\vr_{\mathrm{W4}}$     & S     & $ 1/5$ (a)    & $0.09$    &  \\
        & BS    & $ 0.474$ \cite{TamMultPartEnt}    & $0.434$   & \\ \hline
$\vr_{\mathrm{UPB}}$  & S     & $ 0.87$ \cite{DohertyPRL02}    & $0.83$    & $1.2\cdot 10^{-4}$ \\ \hline
$\vr_{\mathrm{BE3}}$   & S     & $ 0.786^*$ \cite{08HyllusPHDThesis}   & $0.726$   & $2.0\cdot 10^{-4}$ \\
        & BS (AB-C)    & $ 1^*$ \cite{08ClassThreeQubitEntBruss}       & $0.9$     & $1.1\cdot 10^{-3}$ \\ \hline\hline
$\varrho_{\mathrm{H}}(T)$ & S     & $5.61$ \cite{SpinSqueezeIneqToth}      & $5.62$  & $2.1\cdot 10^{-5}$ \\
        & BS (AB-C)& $4.329$ \cite{SpinSqueezeIneqToth}  & $4.33$  & $7.2\cdot 10^{-6}$ \\ \hline
\end{tabular}
\end{center}
\caption{Threshold values for $p$, see Eq.\ (\ref{Eq-WhNoiseP}). With an asterisk we denote exact values from the literature and with (a) we denote bounds obtained via the PPT criterion. The column Decomp.\ contains the parameter threshold up to which we are able to determine S (separable), BS (biseparable), W (W-type) decompositions. The specific methods used to obtain bounds on the threshold values are discussed in the corresponding paragraph of the examples.
\label{Tab-Decomp}
}
\end{table}

\section{Conclusions}
We presented an easy to implement, fast and straight-forward method for finding decompositions of
quantum states with specific SLOCC entanglement properties. For a large variety of examples decompositions
were determined for those parameters, where they are known to exist. For instance, is it possible to
find separable decompositions of Werner states with minimal amount of white noise. Also the threshold
values for biseparability can be reproduced in most cases for three- and four-qubit states. Especially
interesting are the cases where the exact parameter range for the existing decompositions is not known.
It was e.g.\ possible to investigate bound entangled states with our algorithm.
This method of determining specific decompositions of a quantum state is a complementary
tool to entanglement criteria such as entanglement witnesses. Together these tools allow to extensively specify
the entanglement properties of a given quantum state. In the future we would like to understand better the
convergence behavior of the algorithm, especially why optimally decreasing the purity of the quantum
state in each iteration is at the end not necessarily a good strategy.

We would like to thank J. Barreiro, M. Kleinmann, M. Mertz, M. Piani, C. Roos, and A. Streltsov
for  valuable discussions. This project was financially supported by the Austrian Science Fund (FWF):
Y376-N16 (START prize), the EU (Marie Curie CIG 293993/ENFOQI), the BMBF (Chist-Era Project QUASAR)
and the Deutsche Forschungsgemeinschaft (DFG).

\bibliography{QIPLiterature09}

\end{document}